\documentstyle[11pt]{article}
\newcounter{sub}
\newcounter{subeqn}[sub]
\renewcommand{\thesubeqn}{\alph{subeqn}}
\renewcommand{\theequation}{\thesub\thesubeqn}
\begin{document}
\begin{center}
{\Large{\bf  Integrals and static solutions of general relativistic \\Liouville's
equation in post Newtonian approximation}}\vspace{1cm}\\
V. Rezania$^1$ and Y. Sobouti$^{1,2,3}$\\
\end{center}
1. Institute for Advanced Studies in Basic Sciences, P. O. Box
 45195-159,
 Zanjan, Iran.\\
2. Physics Department, Shiraz University, Shiraz, Iran.\\
3. Center for Theoretical Physics and Mathematics, AEOI, Tehran, Iran.\\
e-mail: rezania@sultan.iasbs.ac.ir and sobouti@sultan.iasbs.ac.ir \\ \\
\vspace{3cm}\\
Publication in main journal\\
To appear under section 5, stellar cluster and associations\\
Proofs to be sent to V. Rezania
\newpage
\noindent{\bf Abstract}
The post-Newtonian approximation of general relativistic Liouville's equation
is presented.    Two integrals of it, generalizations of the
classical energy
and angular momentum, are obtained.   Polytropic models are
constructed as an application.
\section{Introduction}
Solutions of general relativistic Liouville's equation ($grl$) in
a prescribed space-time
have been considered by some investigators.
Most authors have sought its solutions as functions of the
constants of motion, generated by Killing vectors of the space-time in
question.   See for example Ehlers (1971), Ray and
Zimmerman (1977),
Mansouri and Rakei (1988), Ellis, Matraverse and Treciokas (1983), Maharaj and
Maartens (1985, 1987), Maharaj (1989), and Dehghani and Rezania (1996).

In applications to self gravitating stars and stellar systems, however, one
should combine Einstein's
field equations and $grl$.   The resulting nonlinear equations can
be solved in certain approximations.
Two such methods are available;
 the {\it post-Newtonian (pn) approximation} and the {\it weak-field
} one.
In this paper we adopt the first approach to study a self gravitating
system imbeded in an otherwise flat space-time.
In sect. 2, we derive the $grl$ in the $pn$ approximation.
In sect. 3 we seek static solutions of
the post-Newtonian Liouville's
equation ($pnl$).     We find two integrals of $pnl$ that are the
$pn$
generalizations of the energy and angular momentum integrals of the
classical Liouville's equation.
Post-Newtonian polytropes, as simultaneous solutions of $pnl$ and Einsteins
equation,
are discussed and calculated in sect.
 4.     Section 5 is devoted to concluding remarks.

The main objective of this paper, however, is to set the stage for the second
paper in this series (Sobouti and Rezania, 1998).     There, we study a class
of
non static oscillatory solutions of $pnl$, different from the conventional $p$
and $g$ modes of the system.      They are associated with oscillations in
the space-time metric, without disturbing the classical equilibrium of the
system.    In this respect they might be akin to the so called {\it
gravitational wave} modes that some authors have advocated to exist in
relativistic systems.     See, for example,
Andersson, Kokkotas and Schutz (1995), Baumgarte and Schmidt (1993),
Detweiler and Lindblom (1983, 1985),
Kojima (1988),
Kokkotas and Schutz (1986, 1992), Leaver (1985),
Leins, Nollert and Soffel (1993),
Lindblom, Mendell and Ipser (1997), and
Nollert and Schmidt (1992).
\section{Liouville's equation in post-Newtonian
approximation}
The one particle  distribution function of a gas of collisionless particles
with identical masse $m$, in the restricted seven dimensional phase space
\stepcounter{sub}
\begin{equation}
P(m):\;\;g_{\mu \nu} U^\mu U^\nu = -1
\end{equation}
satisfies $grl$:
\stepcounter{sub}
\begin{equation}
{\cal L}_Uf = (U^\mu \frac{\partial}{\partial x^\mu} - \Gamma_{\mu\nu}^i U^\mu
U^\nu \frac{\partial}{\partial U^i}) f(x^\mu,U^i) = 0,
\end{equation}
where $(x^\mu,U^i)$ is the set of configuration and velocity coordinates in
 $P(m)$, $f(x^\mu,U^i)$ is a distribution function,
${\cal L}_U$
is Liouville's operator in the $(x^\mu,U^i)$ coordinates, and $\Gamma^i
_{\mu\nu}$ are Christoffel's symbols.
Greek indices run from 0 to 3 and Latin indices from
1 to 3.   We use the convention
$c = 1$ except in numerical calculations of section 4.
  The four-velocity of the particle and
its classical velocity are related as
\stepcounter{sub}
\begin{equation}
U^\mu = U^0 v^\mu;\;\;\;\;  v^\mu = (1, v^i = dx^i/dt),
\end{equation}
where $U^0(x^\mu,v^i)$ is to be determined from Eq. (1).
In the $pn$ approximation, we need an expansion of
${\cal L}_U$ up to order $\bar{v}^4$, where $\bar{v}$ is a typical
Newtonian speed.   To achieve this goal we transform
$(x^\mu,U^i)$ to $(x^\mu,v^i)$.
Liouville's operator transforms as
\stepcounter{sub}
\begin{equation}
{\cal L}_U = U^0 v^\mu (\frac{\partial}{\partial x^\mu} + \frac{\partial v^j}
{\partial x^\mu} \frac{\partial}{\partial v^j}) - \Gamma^i _{\mu\nu}U^{0^2}
v^\mu v^\nu \frac{\partial v^j}{\partial U^i} \frac{\partial}{\partial v^j},
\end{equation}
where  $\partial v^j/ \partial x^\mu$ and $\partial v^j/ \partial U^i$ are
determined
from the inverse of the transformation matrix (see appendix A).
Thus,
\stepcounter{sub}
\stepcounter{subeqn}
\begin{eqnarray}
&&\frac{\partial v^j}{\partial x^\mu} = - \frac{U^0}{2Q} v^j
\frac{\partial g_{\alpha \beta}}{\partial x^\mu} v^\alpha v^\beta,\\
\stepcounter{subeqn}
&&\frac{\partial v^j}{\partial U^i} =
\;\; \frac{1}{Q} v^j (g_{0i} + g_{ik} v^k);\hspace{2cm}\;\;\mbox{for $i
\neq j$},\nonumber\\
&&\\
&&\hspace{.81cm}=-\frac{1}{Q} (U^{0^{-2}} + \sum_{k \neq i} v^k (g_{0 k}
+ g_{kl} v^l));\;\;  \mbox{for $i = j$},\nonumber
\end{eqnarray}
where
\stepcounter{subeqn}
\begin{equation}
Q = U^0 (g_{0 0} + g_{0 l} v^l).
\end{equation}
Using Eqs.(5) in Eq.(4), one finds
\stepcounter{sub}
\stepcounter{subeqn}
\begin{equation}
{\cal L}_U f =U^{0} {\cal L}_v f=0,
\end{equation}
or
\stepcounter{subeqn}
\begin{equation}
{\cal L}_v f(x^\mu,v^i) = 0,
\end{equation}
where
\stepcounter{subeqn}
\begin{eqnarray}
&&{\cal L}_v= v^\mu (\frac{\partial}{\partial x^\mu} - \frac{U^0}{2Q} v^j
\frac{\partial g_{\alpha \beta}}{\partial x^\mu} v^\alpha
v^\beta\frac{\partial}{\partial v^j})
- \Gamma^i _{\mu\nu}v^\mu v^\nu\{\sum_{j\neq i} \frac{1}{Q} v^j (g_{0i}
+ g_{ik} v^k)  \frac{\partial}{\partial v^j}  \nonumber \\
&&\hspace{4cm} - \frac{1}{Q} (U^{0^{-2}} + \sum_{k \neq i} v^k (g_{0 k}
+ g_{kl} v^l)) \frac{\partial}{\partial v^i}   \},
\end{eqnarray}
We note that the post-Newtonian hydrodynamic equations are obtained from
integrations
of Eq. (6a) over the ${\bf v}$-space rather than Eq. (6b) (see appendix B).
We expand
${\cal L}_v$ up to the order $\bar{v}^4$.  For this purpose, we need
expansions of Einstein's field equations, the metric tensor, and the affine
connections up to various orders.
Einstein's field equation with harmonic coordinates
condition, $g^{\mu\nu}\Gamma^{\lambda}_{\mu\nu}=0$, yields (see
Weinberg 1972):
\stepcounter{sub}
\stepcounter{subeqn}
\begin{eqnarray}
&&\nabla^2\; ^2g_{00} = - 8\pi G\; ^0T^{00},\\
&&\nabla^2\; ^4g_{00} = \frac{\partial^2\; ^2g_{00}}{\partial t^2} +
\;^2g_{ij} \frac{\partial^2\; ^2g_{00}}{\partial x^i \partial x^j} -
(\frac{\partial\; ^2g_{00}}{\partial x^i})(\frac{\partial\; ^2g_{00}}
{\partial x^i})\nonumber\\
\stepcounter{subeqn}
&&\hspace{2cm}- 8\pi G (\;^2T^{00} - 2\; ^2g_{00}\; ^0T^{00} +\;
^2T^{ii}),\\
\stepcounter{subeqn}
&&\nabla^2\; ^3g_{i 0} = 16 \pi G\; ^1T^{i0},\\
\stepcounter{subeqn}
&&\nabla^2\; ^2g_{ij} = - 8 \pi G \delta_{ij}\; ^0T^{00}.
\end{eqnarray}
The symbols $^ng_{\mu\nu}$ and $^nT^{\mu\nu}$ denote the $n$th order terms in
${\bar v}$ in
the metric and in the energy-momentum tensors, respectively.    Solutions of
these equations are
\stepcounter{sub}
\stepcounter{subeqn}
\begin{eqnarray}
&&^2g_{00} = - 2 \phi,\\
\stepcounter{subeqn}
&&^2g_{ij} = - 2 \delta_{ij} \phi, \\
\stepcounter{subeqn}
&&^3g_{i 0} = \eta_i,               \\
\stepcounter{subeqn}
&&^4g_{00} = - 2 \phi^2 - 2 \psi,
\end{eqnarray}
where
\stepcounter{sub}
\stepcounter{subeqn}
\begin{eqnarray}
&&\phi({\bf x},t) = -G \int \frac{^0T^{00}({\bf x}',t)}{\vert{\bf x} -
{\bf x}' \vert} d^3 x', \\
\stepcounter{subeqn}
&&\eta^i ({\bf x},t) = -4 G \int \frac{^1T^{i 0} ({\bf x}',t)}{\vert {\bf x} -
{\bf x}' \vert} d^3 x',\\
\stepcounter{subeqn}
&&\psi({\bf x},t) = - \int \frac{d^3 x'}{\vert {\bf x} - {\bf x}' \vert}
(\frac{1}{4\pi} \frac{\partial^2 \phi({\bf x'},t)}{\partial t^2}
+ G\; ^2T^{00}({\bf x}',t)\nonumber\\
\stepcounter{subeqn}
&&\hspace{6cm}+G\;^2T^{ii}({\bf x'}, t)),
\end{eqnarray}
where bold characters denote the three vectors.
Substituting Eqs. (8) and (9) in (6c) gives
\stepcounter{sub}
\begin{eqnarray}
{\cal L}_v &= &{\cal L}^{cl}+{\cal L}^{pn} \nonumber\\
                  & =&\frac{\partial}{\partial t} + v^i
\frac{\partial}{\partial x^i} - \frac{\partial \phi}{\partial x^i}
\frac{\partial}{\partial v^i} \nonumber\\
                &&- [(4\phi + {\bf v}^2) \frac{\partial \phi}{\partial x^i} -
\frac{\partial \phi}{\partial x^j} v^i v^j -  v^i \frac{\partial
\phi}{\partial t} + \frac{\partial \psi}{\partial x^i} \nonumber\\
                  &&+ (\frac{\partial \eta_i}{\partial x^j} -  \frac{\partial
\eta_j}{\partial x^i}) v^j+\frac{\partial\eta_i}{\partial t}]
\frac{\partial}{\partial v^i}
\end{eqnarray}
where ${\cal L}^{cl}$ and ${\cal L}^{pn}$ are the
classical Liouville
operator and its post-Newtonian correction,
respectively.      Equation (6b) for the
distribution function $f(x^\mu,v^i)$ becomes
\stepcounter{sub}
\begin{equation}
({\cal L}^{cl}+{\cal L}^{pn}) f(x^\mu,v^i) = 0.
\end{equation}
The three scalar and vector potentials $\phi,\psi$ and
$\eta\hspace{-.2cm}\eta\hspace{-.2cm}\eta$
can now be given in terms of the distribution function.
    The energy-momentum tensor in terms of $f(x^\mu, U^i)$ is
\stepcounter{sub}
\begin{equation}
T^{\mu\nu}(x^\lambda)=\int \frac{U^\mu U^\nu}{U^0} f(x^\lambda, U^i)
\sqrt{-g}d^3U,
\end{equation}
where $g=det(g_{\mu\nu})$.     For various orders of $T^{\mu\nu}$
 one finds
\stepcounter{sub}
\stepcounter{subeqn}
\begin{eqnarray}
&&^0T^{00}(x^\lambda) = \int  f(x^\lambda,v^i) d^3 v, \\
\stepcounter{subeqn}
&&^2T^{00}(x^\lambda) = \int (\frac{1}{2}v^2 + \phi(x^\lambda))
f(x^\lambda,v^i) d^3 v,\\
\stepcounter{subeqn}
&&^2T^{ij}(x^\lambda)= \int v^i v^j f(x^\lambda,v^i) d^3 v, \\
\stepcounter{subeqn}
&&^1T^{0i}(x^\lambda) = \int v^i f(x^\lambda,v^i) d^3 v.
\end{eqnarray}
Substituting Eqs. (13) in (9) gives
\stepcounter{sub}
\stepcounter{subeqn}
\begin{eqnarray}
&&\phi({\bf x},t) =-G \int \frac{f({\bf x'},t,{\bf v')}}{\vert {\bf
x} - {\bf x'} \vert} d \Gamma',\\
\stepcounter{subeqn}
&&\eta\hspace{-.2cm}\eta\hspace{-.2cm}\eta ({\bf x},t) = -4G \int \frac{{\bf
v'} f({\bf x'},t, {\bf v')}}{\vert {\bf x} - {\bf x'} \vert} d\Gamma'\\
&&\psi({\bf x},t) = \frac{G}{4 \pi} \int \frac{\partial^2
f({\bf x''},t,{\bf v''})/\partial t^2 }{
\vert {\bf x} - {\bf x'} \vert \vert  {\bf
x'} - {\bf x''} \vert} d^3x'd\Gamma'' \nonumber\\
&&\hspace{2cm}- \frac{3}{2}G \int \frac{{\bf v'}^2 f({\bf x'},t, {\bf
v'})}{\vert {\bf x} - {\bf x'} \vert} d \Gamma'\nonumber\\
\stepcounter{subeqn}
&&\hspace{2cm}+ G^2 \int \frac{f({\bf x'},t,{\bf v'}) f({\bf
x''},t,{\bf v''})}{\vert {\bf x} - {\bf x'}
\vert \vert {\bf x'} - {\bf x''} \vert} d \Gamma' d
\Gamma'',
\end{eqnarray}
where $d\Gamma=d^3xd^3v$.      Equations (11) and (14) complete the
$pn$ order of Liouville's equation for self gravitating systems imbeded in
a flat space-time.
\section{Integrals of post-Newtonian Liouville's equation}
In an equilibrium state $f({\bf x}, {\bf v})$ is time-independent.
Macroscopic velocities along with the vector potential
$\eta\hspace{-.2cm}\eta\hspace{-.2cm}\eta$ vanish.       Equations (10)
and (11) reduce to
\stepcounter{sub}
\begin{eqnarray}
&&({\cal L}^{cl}+{\cal L}^{pn})f({\bf x},{\bf v})=[(v^i\frac{\partial}
{\partial x^i}-
\frac{\partial\phi}{\partial x^i}\frac{\partial}{\partial v^i})\nonumber\\
&&\hspace{3cm}-(\frac{\partial\phi}{\partial x^i}(4\phi+v^2)-
\frac{\partial\phi}{\partial x^j}v^iv^j+\frac{\partial\psi}{\partial x^i})
\frac{\partial}{\partial v^i}]f=0,
\end{eqnarray}
One easily verifies that the following, a generalization of the classical
energy integral, is a solution of Eq. (15)
\stepcounter{sub}
\begin{eqnarray}
&&E=\frac{1}{2}v^2+\phi+2\phi^2+\psi + const.
\end{eqnarray}
Furthermore, if $\phi({\bf x})$ and $\psi({\bf x})$ are spherically symmetric,
which actually is the
case for an isolated system in an asymptotically flat space-time,
the following generalization of angular momentum are also integrals of
Eq. (15)
\stepcounter{sub}
\begin{eqnarray}
&& l_i=\varepsilon_{ijk}x^jv^kexp(-\phi),
\end{eqnarray}
where $\varepsilon_{ijk}$ is the Levi-Cevita symbol.      Static distribution
functions maybe constructed as functions of $E$ and even functions of
$l_i$.       The reason for restriction
to even functions of $l^{pn}_i$ is to insure the vanishing of $\eta^i$,
the condition for validity of Eq. (15).
\section{Polytropes in post-Newtonian approximation}
As in classical polytropes we
consider the distribution
function for a polytrope of index $n$ as
\stepcounter{sub}
\begin{eqnarray}
&&F_n(E)=\frac{\alpha_n}{4\pi\sqrt{2}}(-E)^{n-3/2};\;\;
\mbox{for}\;\; E< 0, \nonumber\\
&&\hspace{1.3cm}=0\hspace{3.00cm}\mbox{for}\;\; E> 0,
\end{eqnarray}
where $\alpha_n$ is a constant.
By Eqs. (13) the corresponding orders of the energy-momentum tensor are
\stepcounter{sub}
\stepcounter{subeqn}
\begin{eqnarray}
&&^0T^{00}_n=\alpha_n\beta_n(-U)^n,\\
\stepcounter{subeqn}
&&^2T^{00}_n=\alpha_n\beta_n\phi (-U)^n +\alpha_n\gamma_n
(-U)^{n+1},\\ \stepcounter{subeqn}
&&^2T^{ii}_n\;\;=\delta_{ij}\;^2T^{ij}=2\alpha_n\gamma_n (-U)^{n+1},\\
\stepcounter{subeqn}
&&^1T^{0i}_n=0,
\end{eqnarray}
where
\stepcounter{sub}
\begin{eqnarray}
&&\beta_n=\int^1_0(1-\eta)^{n-3/2}\eta^{1/2}d\eta=\Gamma(3/2)\Gamma(n-1
/2)/\Gamma(n+1),\\
\stepcounter{sub}
&&\gamma_n=\int^1_0(1-\eta)^{n-3/2}\eta^{3/2}d\eta=\Gamma(5/2)\Gamma(n-1/
2)/\Gamma(n+2),
\end{eqnarray}
and $U=\phi+2\phi^2+\psi$ is the gravitational potential in $pn$ order.
It will be chosen zero at the
surface of the stellar configuration.     With this choice, the escape velocity
$v_e=\sqrt{-2U}$ will mean escape to the boundary of the system rather
than to infinity.   Einstein's equations, Eqs. (7), (8) and (9), lead to
\stepcounter{sub}
\begin{eqnarray}
&&\nabla^2\phi=4\pi G\; ^0T^{00}=
4\pi G\alpha_n\beta_n(-U)^n,\\
\stepcounter{sub}
&&\nabla^2\psi=4\pi G (^2T^{00}+^2T^{ii})
=4\pi G\alpha_n\beta_n\phi (-U)^n\nonumber\\
&&\hspace{1cm}+12\pi G\alpha_n\gamma_n(-U)^{n+1}.
\end{eqnarray}
Expanding $(-U)^n$ as
\stepcounter{sub}
\begin{equation}
(-U)^n=(-\phi)^n[1+n(2\phi+\frac{\psi}{\phi})+\cdots],
\end{equation}
ans Substituting it in Eqs. (22) and (23) gives
\stepcounter{sub}
\begin{eqnarray}
&&\nabla^2\phi=4\pi
G\alpha_n\beta_n[(-\phi)^n-2n(-\phi)^{n+1}-n(-\phi)^{n-1}\psi+\cdots],\\
\stepcounter{sub}
&&\nabla^2\psi=4\pi
G\alpha_n\beta_n
\{(3\frac{\gamma_n}{\beta_n}-1) (-\phi)^{n+1} \nonumber\\
&&\hspace{3cm}- [3(n+1)\frac{\gamma_n}{\beta_n}-n] [2(-\phi)^{n+2}+
(-\phi)^n\psi]+\cdots\},
\end{eqnarray}
These equations can be solved numerically by an iterative scheme.    We
introduce three
dimesionless quantities
\stepcounter{sub}
\stepcounter{subeqn}
\begin{eqnarray}
&&-\phi\equiv \lambda \theta,\\
\stepcounter{subeqn}
&&-\psi\equiv \lambda^2 \Theta,\\
\stepcounter{subeqn}
&&\hspace{.35cm}r\equiv a\zeta,
\end{eqnarray}
where, in terms of $\rho_c$, the central density,
$\lambda=(\rho_c/\alpha_n\beta_n)^{1/n}$ and
$a^{-2}=4\pi G\rho_c/\lambda$.      Equations
(25) and (26) in various iteration orders reduce to
\stepcounter{sub}
\stepcounter{subeqn}
\begin{eqnarray}
&&\nabla_{\zeta}^2\theta_o + \theta_o^n=0,\\
\stepcounter{subeqn}
&&\nabla_{\zeta}^2\Theta_o +(3\frac{\gamma_n}{\beta_n}-1) \theta_o^{n+1}=0,\\
\stepcounter{subeqn}
&&\nabla_{\zeta}^2\theta_1+\theta_1^n = qn(2\theta_o^{n+1} - \theta_o^{n-1}
\Theta_o),\\
&&\nabla_{\zeta}^2\Theta_1+(3\frac{\gamma_n}{\beta_n}-1) \theta_1^{n+1}=
\nonumber\\
\stepcounter{subeqn}
&&\hspace{2cm}q [3(n+1)\frac{\gamma_n}{\beta_n}-n] (2\theta_o^{n+2}-
\theta_o^n\Theta_o),
\end{eqnarray}
where
$\nabla_{\zeta}^2= \frac{1}{\zeta^2}
\frac{d}{d\zeta}(\zeta^2\frac{d}{d\zeta})$.
The subscripts 0 and 1 refer to orders of the iteration.   The
dimensionless parameter $q$ is defined as
\stepcounter{sub}
\begin{equation}
q=\frac{4\pi G \rho_c a^2}{c^2}=\frac{R_s}{R} \frac{1}{2\zeta_1\mid\theta_o'
(\zeta_1)\mid},
\end{equation}
where $R_s$ is the Schwarzschild radius, $R=a\zeta_1$ is the radius of system,
$\zeta_1$ is defined by $\theta_o(\zeta_1)=0$ and $c$ is the light speed.
The order of magnitude of $q$ varies from $10^{-5}$  for white dwarfs to
$10^{-1}$ for neutron stars.       For future reference, let us also note that
\stepcounter{sub}
\begin{equation}
-U=\lambda[\theta_1 + q (\Theta_1 - 2\theta_1^2)].
\end{equation}
We use a forth-order Runge-Kutta method to find
numerical solutions of the four coupled nonlinear differential Eqs. (28).
At the center we adopt
\stepcounter{sub}
\begin{equation}
\theta_a(0)=1;\;\;\;\theta'_a(0)=\frac{d\theta_a}{d\zeta}\mid_0=0;\;\;
\;\;\; a=0,1.
\end{equation}
In tables 1 and 2, we summarize the numerical results for the Newtonian and
post-Newtonian polytropes for different polytropic indices and $q$ values.
The $pn$ corrections tend to reduce the radius of the polytrope.
The
higher the polytropic index the smaller this radius.      The same is true,
of course, for higher values of $q$.\\
\section{Concluding remarks}
As we discussed in section 1, some authors is debated to exist a new modes of
oscillations in relativistic stellar systems.      They believed that these
modes generated by the perturbations in space time metric and no have analogue
in Newtonian systems.       They used  the general relativistic hydrodynamics
to distinguish them.
Although this way is routine but one needs to assume some thermodynamic
concepts
that may be fault in relativistic regime.    Hence, to reject these conceptual
problems, we choosed general relativistic Liouville's equation that is the
purely dynamical theory.       The combination Liouville and Einstein equations
enable one to study the behavior of relativistic systems without ambiguity.
Therefore, in this paper, we used the $pn$ approximation to obtain the
Einstein-Liouville equation for a relativistic self gravitating stellar system.
    We found two integrals,
generalization of the classical energy and angular momentum, that are
satisfying $pnl$ in equilibrium state.      These solutions enable one to
construct an equilibrium model for the system in $pn$ approximation.
Polytropic models, the most familiar stellar models, are constructed in $pn$
approximation.     In tables 1 and 2, we compared these models with its
Newtonian correspondence.     The $pn$ corrections tend to reduce the radius of
polytrope.      The higher the polytropic index the smaller this radius.
We introduced a parameter $q$, Eq. (29), to enter the effect of central density
of system in calculations.      Increasing values of $q$ reduce the
radius of system.       In the second paper (Sobouti and Rezania 1998), we
study time-dependent oscillations of a relativistic system in $pn$
approximation.

\newpage
\setcounter{sub}{0}
\setcounter{subeqn}{0}
\renewcommand{\theequation}{A.\thesub\thesubeqn}
\noindent {\large{\bf Appendix A: Derivation of Eqs. (5)
}}\\ Consider a general coordinate transformation $(X, U)$ to
$(Y, V)$.   The corresponding partial derivatives transform as
\[\left( \begin{array}{c}
\partial/\partial X\\ \partial/\partial U
          \end{array} \right)\;= M
\left( \begin{array}{c}
\partial/\partial Y\\ \partial/\partial V
          \end{array} \right)\;,\]

\stepcounter{sub}
\begin{equation}
\hspace{5.9cm}=
\left( \begin{array}{cc}
 \partial Y/\partial X &\partial V/\partial
X\\
\partial Y/\partial U &\partial V/\partial U
            \end{array}\right)
\left( \begin{array}{c}
\partial/\partial Y\\\partial/\partial V
          \end{array} \right) ,
\end{equation}
where $M$ is the $7\times 7$ Jacobian matrix of transformation.   Setting
$X=Y=x^{\mu}$, $V=v^i$ and $U=U^i$ for our problem, one finds
\stepcounter{sub}
\stepcounter{subeqn}
\begin{equation}
M=\left( \begin{array}{cc}
\partial x^{\mu}/\partial x^{\nu}&\partial v^i/\partial
x^{\nu}\\
\partial x^{\mu}/\partial U^j&\partial v^i/\partial U^j
\end{array}\right),
\end{equation}
and its inverse
\stepcounter{subeqn}
\begin{equation}
M^{-1}=\left( \begin{array}{cc}
\partial x^{\mu}/\partial x^{\nu}&\partial U^i/\partial
x^{\nu}\\
\partial x^{\mu}/\partial v^j&\partial U^i/\partial v^j
\end{array}\right).
\end{equation}
One easily finds
\stepcounter{sub}
\stepcounter{subeqn}
\begin{eqnarray}
&&\partial x^{\mu}/\partial x^{\nu}=\delta_{\mu\nu};\;\;\;\;\;\;\;
\partial x^{\mu}/\partial v^j=0,\\
\stepcounter{subeqn}
&&\partial U^i/\partial x^{\nu}=v^i\partial U^0/\partial
x^{\nu}
=\frac{{U^0}^3v^i}{2}\frac{\partial g_{\alpha\beta}}{\partial
x^{\nu}}v^{\alpha}v^{\beta},\\
\stepcounter{subeqn}
&&\partial U^i/\partial v^j=U^0\delta_{ij}+v^i \partial
U^0/\partial v^j
=U^0\delta_{ij}+{U^0}^3v^ig_{j\beta}v^{\beta}.
\end{eqnarray}
Inserting the latter in $M^{-1}$ and inverting the result one arrives at $M$
from which Eqs. (5) can be read out.\\
\newpage
\setcounter{sub}{0}
\renewcommand{\theequation}{B.\thesub\thesubeqn}
\noindent {\large{\bf Appendix B: Post-Newtonian hydrodynamics}}\\
Mathematical manipulations in the development of this work has been tasking.
To ensure that no error has crept in the course of calculations we have tried
to infer the post-Newtonian hydrodynamical equations from the post-Newtonian
Liouville equation derived earlier.        From Eq. (6a) one has
\stepcounter{sub}
\begin{eqnarray}
{\cal L}_U^{pn}f&&=U^0({\cal L}^{cl}+{\cal L}^{pn})f\nonumber\\
&&=[(1-\phi+\frac{1}{2}{\bf v}^2){\cal L}^{cl}+{\cal L}^{pn}]f,
\end{eqnarray}
where ${\cal L}^{cl}$ and ${\cal L}^{pn}$ are given by Eq. (10).
    We integrate ${\cal L}_U^{pn}f$ over the ${\bf v}$-space:
\stepcounter{sub}
\begin{equation}
\int {\cal L}_U^{pn}fd^3v=\int [(1-\phi+\frac{1}{2}{\bf v}^2) {\cal
L}^{cl}+{\cal L}^{pn}]fd^3v.
\end{equation}
Using Eqs. (12) and (13), one finds the continuity equation
\stepcounter{sub}
\begin{eqnarray}
&&\frac{\partial}{\partial
t}(\;^0T^{00}+\;^2T^{00})+\frac{\partial}{\partial x^j}(\; ^1T^{0j}
+\; ^3T^{0j})-\;^0T^{00}\frac{\partial \phi}{\partial t} =0,\nonumber\\
\end{eqnarray}
which is the $pn$ expansion of the continuity equation
\stepcounter{sub}
\begin{equation}
T^{0\nu}_{\;\;\;;\nu}=0,
\end{equation}
Next,
we multiply ${\cal L}_U^{pn}f$ by $v^i$ and integrate over the ${\bf
v}$-space:
\stepcounter{sub}
\begin{equation}
\int v^i {\cal L}_U^{pn}fd^3v=\int v^i[(1-\phi+\frac{1}{2}{\bf v}^2) {\cal
L}^{cl}+{\cal L}^{pn}]fd^3v.
\end{equation}
After some calculations one finds
\stepcounter{sub}
\begin{eqnarray}
&&\frac{\partial}{\partial
t}(\;^1T^{0i}+\;^3T^{0i})+\frac{\partial}{\partial x^j}(\; ^2T^{ij}
+\; ^4T^{ij})\nonumber\\
&&\;\;\;+\; ^0T^{00}[\frac{\partial}{\partial x^i}(\phi+2\phi^2+\psi)+
\frac{\partial
\eta_i}{\partial t}]+\; ^2T^{00}\frac{\partial
\phi}{\partial x^i}\nonumber\\
&&\;\;\;+\; ^1T^{0j}(\frac{\partial \eta_i}{\partial x^j}-\frac{\partial
\eta_j}{\partial x^i}-4\delta_{ij}\frac{\partial \phi}{\partial t})+ \;
^2T^{jk}
(\delta_{jk}\frac{\partial \phi}{\partial x^i}-4\delta_{ik}\frac{\partial \phi}
{\partial x^j})=0.
\end{eqnarray}
The latter is the correct $pn$ expansion of
\stepcounter{sub}
\begin{equation}
T^{i\nu}_{\;\;\;;\nu}=0;\;\;\;\;i=1,2,3.
\end{equation}
See Weinberg 1972, QED.
  \vspace{2cm}\\

{\Large{\bf References:}}\\
\begin{itemize}
\item[ ] Andersson N., Kokkotas K. D., Schutz B. F., 1995, M. N. R. A. S.,
{\bf 274}, 1039
\item[ ] Baumgarte, T. W., Schmidt, B. G., 1993, Class. Quantum. Grav.,
{\bf 10}, 2067
\item[ ] Dehghani M. H., Rezania V., 1996, Astron. Astrophys., {\bf 305}, 379
\item[ ] Detweiler S. L., Lindblom L., 1985, Ap. J., {\bf 292}, 12
\item[ ] Ehlers J., 1977, in: Sachs R. K. (ed.) Proceedings of the international
summer school of
\indent Physics ``Enrico Fermi'', Course 47
\item[ ] Ellis G. R., Matraverse D. R., Treciokas R., 1983, Ann. Phys. {\bf
150}, 455
\item[ ] Ellis G. R., Matraverse D. R., Treciokas R., 1983, Ann. Phys. {\bf
150}, 487
\item[ ] Ellis G. R., Matraverse D. R., Treciokas R., 1983, Gen. Rel. Grav.
{\bf 15}, 931
\item[ ] Kojima, Y., 1988, Prog. Theor. Phys.,
{\bf 79}, 665
\item[ ] Kokkotas K. D., Schutz B. F., 1986, Gen. Rel. Grav., {\bf 18},
913
\item[ ] Kokkotas K. D., Schutz B. F., 1992, M. N. R. A. S., {\bf
255}, 119
\item[ ] Leaver, E. W., 1985, Proc. R. Soc. London A,
{\bf 402}, 285
\item[ ] Leins, M., Nollert, H. P., Soffel, M. H., 1993, Phys. Rev. D,
{\bf 48}, 3467
\item[ ] Lindblom L., Detweiler S. L., 1983, Ap. J. Suppl., {\bf 53}, 73
\item[ ] Lindblom, L., Mendell, G., Ipser, J. R., 1997, Phys. Rev. D,
{\bf 56}, 2118
\item[ ] Maartens R., Maharaj S. D., 1985, J. Math. Phys. {\bf 26}, 2869
\item[ ] Maharaj S. D., Maartens R., 1987, Gen. Rel. Grav. {\bf 19}, 499
\item[ ] Maharaj S. D., 1989, Nouvo Cimento 163B, No. 4, 413
\item[ ] Mansouri A., Rakei A., 1988, Class. Quantum. Grav. {\bf 5}, 321
\item[ ] Nollert, H. -P., Schmidt, B. G., 1992, Phys. Rev. D,
{\bf 45}, 2617
\item[ ] Ray J. R., Zimmerman J. C., 1977, Nouvo Cimento 42B, No. 2, 183
\item[ ] Sobouti Y., Rezania V., 1998, Astron. Astrophys., (Submitted to)
\item[ ] Weinberg S., 1972, Gravitation and Cosmology. John Wiley \& Sons, New
york
\end{itemize}
\newpage
\noindent Table 1.  A comparison of the Newtonian and post-Newtonian polytropes
at certain selected radii for $n$=1, 2 and 3 and
different values $q$.\vspace{1cm}\\
Table 2.   Same as Table 1. $n$=4 and 4.5.
\newpage
\begin{center}
Table 1.\vspace{.2cm}\\
\begin{tabular}{|c|c|c|c|c|c|}\hline
n &Polytropic & Newtonian   &\multicolumn{3}{c|}{$pn$ polytrope,
$\theta+q(\Theta-2\theta^2)$}\\\cline{4-6}
  &radius, $\zeta$ &polytrope, $\theta$& $q=10^{-5}$ & $q=10^{-3}$ &
$q=10^{-1}$ \\\hline   & 0.0000000  & 1.00000 & 1.00000 & 1.00000 & 1.00000\\
  & 1.0000000  & 0.84145 & 0.84145 & 0.84143 & 0.83936\\
  & 2.0000000  & 0.45458 & 0.45458 & 0.45433 & 0.42949\\
1 & 2.9233000  & 0.07408 & 0.07407 & 0.07334 &  0.00000    \\
  & 3.1388500  & 0.00087 & 0.00086 & 0.00000 &         \\
  & 3.1415500  & 0.00001 & 0.00000 &         &      \\
  & 3.1415930  & 0.00000 &         &         &      \\\hline
  & 0.0000000  & 1.00000 & 1.00000 & 1.00000 & 1.00000\\
  & 1.0000000  & 0.84868 & 0.84868 & 0.84863 & 0.84394\\
  & 2.0000000  & 0.52989 & 0.52988 & 0.52945 & 0.48609\\
  & 3.0000000  & 0.24188 & 0.24187 & 0.24031 & 0.13289\\
2 & 3.4737000  & 0.13904 & 0.13902 & 0.13770 &  0.00000     \\
  & 4.3394800  & 0.00171 & 0.00169 & 0.00000 &      \\
  & 4.3527000  & 0.00002 & 0.00000 &         &       \\
  & 4.3529000  & 0.00000 &         &         &       \\\hline
  & 0.0000000  & 1.00000 & 1.00000 & 1.00000 & 1.00000\\
  & 1.0000000  & 0.85480 & 0.85480 & 0.85473 & 0.84773\\
  & 2.0000000  & 0.58284 & 0.58283 & 0.58230 & 0.52894\\
  & 3.0000000  & 0.35939 & 0.35938 & 0.35824 & 0.24016\\
  & 4.0000000  & 0.20927 & 0.20925 & 0.20764 & 0.03226\\
3 & 4.1939500  & 0.18690 & 0.18688 & 0.18520 &  0.00000     \\
  & 6.8435000  & 0.00228 & 0.00225 & 0.00000 &              \\
  & 6.8963000  & 0.00002 & 0.00000 &         &               \\
  & 6.8967000  & 0.00000 &         &         &              \\\hline
\end{tabular}
\end{center}
\newpage
\begin{center}
Table 2.\vspace{.2cm}\\
\begin{tabular}{|c|c|c|c|c|c|}\hline
n &Polytropic & Newtonian   &\multicolumn{3}{c|}{$pn$ polytrope,
$\theta+q(\Theta-2\theta^2)$}\\\cline{4-6}
  &radius, $\zeta$ &polytrope, $\theta$& $q=10^{-5}$ & $q=10^{-3}$ &
$q=10^{-1}$ \\\hline   & 0.0000000  & 1.00000 & 1.00000 & 1.00000 & 1.00000\\
  & 2.0000000  & 0.62294 & 0.62293 & 0.62235 & 0.56326\\
  & 4.0000000  & 0.31804 & 0.31802 & 0.31645 & 0.14194\\
  & 5.1541000  & 0.22574 & 0.22572 & 0.22383 &  0.00000    \\
  & 8.0000000  & 0.10450 & 0.10448 & 0.10221 &         \\
4 &12.0000000  & 0.02972 & 0.02970 & 0.02716 &         \\
  &14.0000000  & 0.00833 & 0.00830 & 0.00570 &         \\
  &14.6468000  & 0.00265 & 0.00262 & 0.00000 &         \\
  &14.9680000  & 0.00003 & 0.00000 &         &      \\
  &14.9713400  & 0.00000 &         &         &      \\\hline
  & 0.0000000  & 1.00000 & 1.00000 & 1.00000 & 1.00000\\
  & 2.0000000  & 0.63965 & 0.63964 & 0.63905 & 0.57857\\
  & 4.0000000  & 0.36053 & 0.36651 & 0.35897 & 0.18628\\
  & 5.7468600  & 0.24334 & 0.24332 & 0.24135 & 0.00000\\
4.5 & 8.0000000& 0.16173 & 0.16171 & 0.15946 &     \\
  &12.0000000  & 0.09015 & 0.09013 & 0.08766 &      \\
  &16.0000000  & 0.05402 & 0.05399 & 0.05141 &       \\
  &20.0000000  & 0.03230 & 0.03227 & 0.02962 & \\
  &24.0000000  & 0.01782 & 0.01779 & 0.01510 & \\
  &28.0000000  & 0.00747 & 0.00744 & 0.00472 & \\
  &30.2689200  & 0.00282 & 0.00279 & 0.00000 &\\
  &31.7792300  & 0.00004 & 0.00000 &         &    \\
  &31.7878400  & 0.00000 &         &         &  \\\hline
\end{tabular}
\end{center}
\end{document}